# Nanoscale Magnetic Domain Memory

Karine Chesnel

Additional information is available at the end of the chapter

> *"Magnets are a bit like humans.*
> *Not only they attract each other, they also have the capability to remember."*

**Abstract**

Magnetic domain memory (MDM) is the ability exhibited by certain magnetic materials to reproduce the exact same nanoscale magnetic domain pattern, even after it has been completely erased by an external magnetic field. In this chapter, we review the various circumstances under which this unusual phenomenon occurs. We explain how partial MDM was first observed in rough Co/Pt multilayers with perpendicular magnetization as a result of structural defects. We then show how 100 % MDM was achieved, even in smooth ferromagnetic films, by coupling Co/Pd multilayers to an antiferromagnetic IrMn template via exchange interactions. We describe how high MDM, extending throughout nearly the entirety of the magnetization process, is obtained when zero-field-cooling the material below its blocking temperature where exchange couplings occur. We also review the persistence of MDM through field cycling and while warming the material all the way up to the blocking temperature. Additionally, we discuss the spatial dependence of MDM, highlighting intriguing oscillatory behaviors suggesting magnetic correlations and rotational symmetries at the nanoscopic scales. Finally, we review the dependence of MDM on cooling conditions, revealing how MDM can be fully controlled, turned on and off, by adjusting the magnitude of the cooling field.

## 1. Introduction to the principles of MDM

### 1.1. Magnetic domain patterns in ferromagnetic films

Ferromagnetic materials are typically composed of magnetic domains. A magnetic domain is a region where the magnetic moments carried by individual atoms, i.e., the atomic spins, align in the same direction due to exchange couplings, as illustrated in **Figure 1a**. Rather than forming one giant macroscopic magnetic domain, the material often breaks down into a multitude of microscopic domains of different orientations, as illustrated in **Figure 1b**. At the delimitation between one magnetic domain and a neighboring one, a coherent rotation of the atomic spins occurs. The delimiting region, called domain wall, may be small in respect to the domain size. Domain sizes typically range from 1 to 100 μm in bulk ferromagnetic materials, and from 100 nm to 1 μm in thin ferromagnetic films [1–3].

In a ferromagnetic material, the microscopic magnetic domains of various magnetization directions arrange in specific ways that minimize the competing magnetic energies present in the system. The three dimensional sum of the magnetic moments carried by the individual domains produces, at the macroscopic scale, a net magnetization *M*. The magnitude and the direction of the net magnetization depends on how the magnetic moments of the individual domains are distributed throughout the material. While *M* may be uniquely set by applying an external magnetic field *H*, following a specific magnetization procedure, the associated microscopic magnetic domain pattern, or topology, is usually not unique. The formation of magnetic domain topologies and their correlation with the magnetic history of the material is still today a vast field of research to be explored and understood.

Thin ferromagnetic films have been extensively studied during the past decades, in particular because of the variety of magnetic domain patterns they exhibit [4–8]. Depending on the composition, the crystallographic structure and the thickness of the film, the magnetization may point in-plane or out-of-plane. The magnetic domains may take various shapes, from round bubbles to elongated, almost infinite, stripes. Some thin film materials exhibit complex formations such as vortices [9–11] and skyrmions [12, 13], illustrated in **Figure 1c, d**, which have instigated a renewed interest in the recent years because of their potential applications in magnetic memory technologies and spintronics [14, 15].

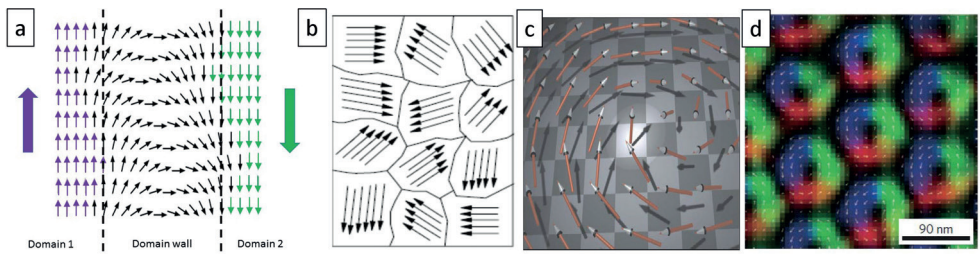

**Figure 1.** (a) Schematic illustrating the formation of magnetic domains by alignment of atomic spins. The schematic shows a domain wall, formed by the coherent rotation of spins. (b) Schematic of a collection of magnetic domains in a ferromagnetic material. (c) Schematic of a magnetic vortex core [10]. (d) Lorentz microscopy image of a skyrmion lattice in FeCoSi. Extracted from Fert et al. [14].

### 1.2. The case of thin films with perpendicular magnetic anisotropy

Thin ferromagnetic films with perpendicular magnetic anisotropy (PMA) [16] exhibit a particularly rich set of magnetic domain patterns. In PMA films, the domain magnetization points mostly out-of-plane, thus producing high demagnetization fields and leading to the formation of regular domain patterns and magnetic textures [17–21]. PMA films have attracted an accrued attention in the years 2000–2010s because the perpendicular magnetization has enabled significant domain size reductions, in comparison to the in-plane magnetic films, thus benefiting the magnetic recording industry. State-of-the-art ultra-high density magnetic recording technologies utilizes granular PMA thin film media, where magnetic domain sizes are as small as 20–50 nm [22].

In smooth PMA films, magnetic domains take a variety of shapes and they form patterns of various topologies. Such variety of shapes and topologies is for example observed in $[Co/Pt]_N$ multilayers where the thickness of Co typically varies between 5 and 50 Å, the thickness of the Pt layer is around 7 Å, and the number of repeat around N = 50. In these multilayers, the PMA is primarily achieved by exploiting the surface anisotropy created by the layering between the Co and the Pt layers, and the high number of repeats, as well as the magneto-crystalline anisotropy produced by the crystallographic texturing [23–26].

The magnetic domain topologies observed in these Co/Pt multilayers vary from bubble domain patterns to pure maze patterns formed of long interlaced striped domains, as shown in **Figure 2**. Both patterns in **Figure 2** are observed at remanence, where the external field is $H = 0$. At that point, the net magnetization is $M \approx 0$, so the area covered by the domains of a given direction nearly equals the area covered by the domains in the opposite direction. Both the bubble pattern and the maze pattern, as well as a wide array of intermediate patterns, can be observed at remanence in the same multilayer. The topology of the remanent pattern mainly depends on the magnetic history of the film. In particular, it has been shown that the remanent magnetic topology drastically change when the magnitude $H_m$ of the magnetic field applied perpendicular to the film changes. If the applied field is saturating, the film form a maze-like pattern at remanence. However, if $H_m$ is set to a specific value slightly below the saturation point, a bubble pattern is achieved and the domain density is maximized [27]. Each time the external magnetic field is cycled, the magnetic domain pattern is erased. When the field is released back to $H = 0$, the new remanent domain pattern is, in this material, different from the previous one.

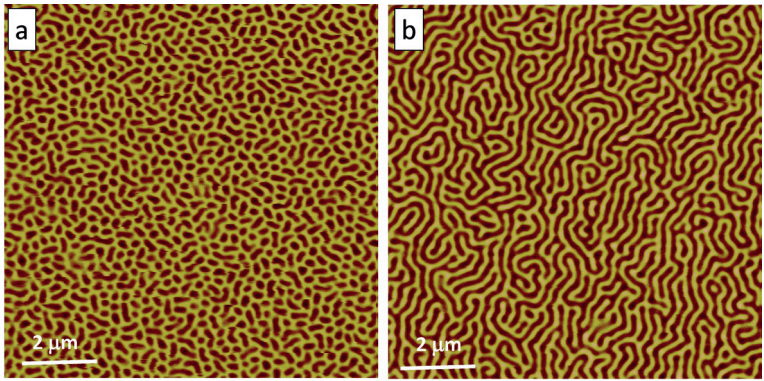

**Figure 2.** Magnetic Force Microscopy (MFM) images of magnetic domains in a $[Co(31\ Å)/Pt(7\ Å)]_{50}$ multilayer with perpendicular magnetic anisotropy. These domains patterns are measured at remanence ($H = 0$). The two colors represent opposite magnetization directions, pointing out-of-plane and into the plane. Image size is 10 × 10 μm. (a) Maze pattern forming after applying $H$ = 2000 Oe; (b) bubble pattern forming after applying $H$ = 9000 Oe. The saturation point is $H$ = 11,400 Oe for this film. Extracted from Chesnel et al. [27].

**1.3. Magnetic domain memory**

Magnetic domain memory (MDM) is the tendency for magnetic domains to retrieve the same exact same pattern after the pattern has been erased by a saturating magnetic field. When MDM occurs, not only the magnetic domains retrieve the same type of shape and size, but their distribution in space, or topology, is similar. MDM may be total, in which case the domain topology is identical; or MDM may be partial, in which case the domain pattern is similar, with some topological differences.

Ferromagnetic films usually do not exhibit any MDM. Like in the Co/Pt multilayers previously mentioned, the domain pattern at a given field point $H$ generally does not repeat after the field has been cycled [28]. The magnetic domains may always take the form of small bubbles or of elongated stripes of specific size at that point in field, but their spatial distribution will be completely different after each cycle. This is well illustrated in **Figure 3**, which shows magnetic domain patterns in Co/Pt multilayers measured at remanence before and after applying a minor magnetization cycle. The initial pattern exhibits elongated striped domains of opposite up / down magnetization with close to a 50% up: 50% down coverage. At the highest magnetic field value, which is not quite saturating, the domains of reverse magnetization have shrunk down to small sparse bubbles. Upon return to zero field, the film exhibits magnetic stripes again but these stripes are shorter and their spatial distribution, or topology, is completely different than in the initial image. In this case, the film does not exhibit any MDM.

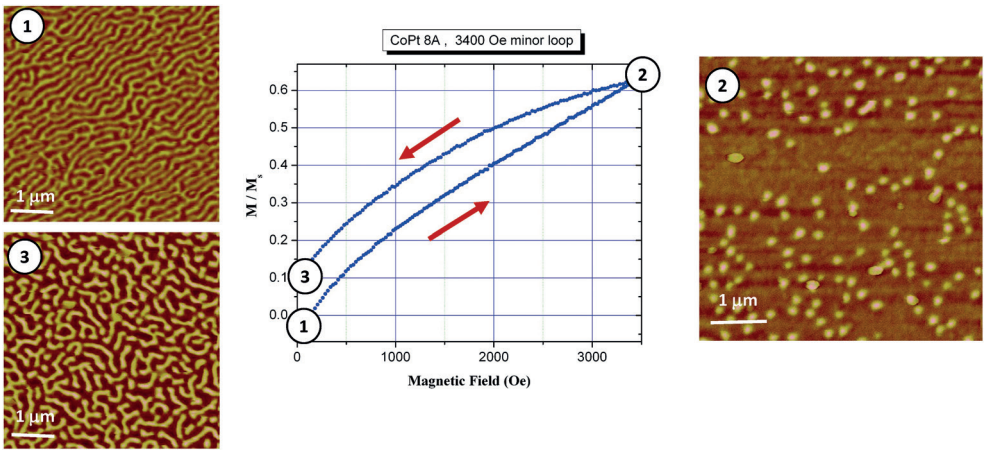

**Figure 3.** Evolution of the magnetic domain pattern in a [Co(8 Å)/Pt(7 Å)]$_{50}$ multilayer while cycling the magnetic field $H$ through half a minor loop. Plotted is the net magnetization $M$ ($H$). The MFM images are 5 × 5 μm and are all collected at the same location of the film. (1) At remanence ($H = 0$), before applying the field; (2) at $H = 3400$ Oe, domains have shrunk down to small bubbles; (3) back to remanence ($H = 0$), after applying the field. Extracted from Westover et al. [28].

MDM has been so far only observed in specific thin ferromagnetic films that have specific structural or magnetic properties. As explained in the following sections, MDM has first been discovered in rough Co/Pt films, due to the presence of defects, acting as pinning sites for the domain nucleation. In that case, the observed defect-induced MDM is partial and only occurs in the nucleation phase of the magnetization cycle. We will see how MDM can however be maximized by exploiting magnetic exchange between the ferromagnetic (F) layer and an antiferromagnetic (AF) layer. This has been successfully achieved in [Co/Pd] IrMn films where F Co/Pd multilayers are sandwiched in between AF IrMn layers. The film is field-cooled (FC) down below its blocking temperature, to allow exchange couplings to occur. Under these conditions, the magnetic domain pattern imprinted in the IrMn layer plays the role of a magnetic template. In this case, the observed exchange-bias induced MDM reaches 100% throughout a large portion of the magnetization process.

## 2. Probing MDM

### 2.1. Real space imaging

Magnetic domain patterns may be directly measured via Magnetic Force Microscopy (MFM), as seen in **Figures 2** and **3**. The MFM technique allows the visualization of magnetic domains by probing the out-of-plane magnetic stray fields emanating from the surface of the film. With a spatial resolution down to about 20–25 nm, MFM allows the detection of individual magnetic domains in thin PMA ferromagnetic films, as these domains are typically 50–200 nm wide.

The investigation of MDM may be possible via MFM under certain experimental conditions. In 2003, Kappenberger et al. [29] showed in their study of CoO/[Co/Pt] multilayers via MFM that the specific magnetic domain pattern initially observed in a particular region of the film was fully recovered after magnetically saturating the material, as seen in **Figure 4**. In this experiment, the material was zero-field-cooled (ZFC) down to 7.5 K and then imaged at 7.5 K while applying a large magnetic field up to 7 T. The initial striped domain pattern observed in the Co/Pt layer immediately after ZFC was completely erased under application of magnetic field, but it was fully retrieved when the magnetic field was decreased back to near coercive point $H_c$ on the descending branch of the magnetization loop. At $H_c$, the net magnetization is near zero, like it was during ZFC.

Such experiment requires very sophisticated MFM instrumentation, with cryogenic and in-situ magnetic field capabilities. Only very few MFM instruments in the world allow such extreme temperature and field environments. Because of the difficulty to measure MFM images at low temperature and under high in-situ magnetic field, the study of MDM via MFM is somewhat limited in practice. The magnetization of the MFM tip may reverse while the in-situ magnetic field is being varied. Also the maximum possible value for the magnetic field may be smaller than the value required to saturate the material, making the collection of MFM images throughout the entire magnetization loop impossible. Lastly, MFM images are typically micrometric, covering a few microns of the film, and a handful of magnetic domains, thus providing a localized view only.

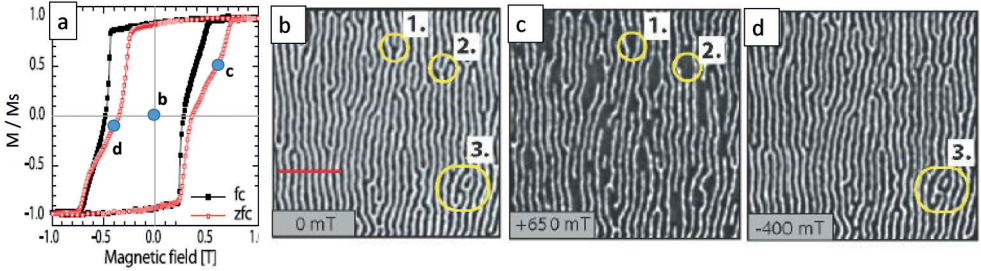

**Figure 4.** Evolution of the magnetic domain pattern in an exchange-biased CoO/ [Co/Pt] multilayer. The MFM images are 5 × 5 μm. (a) Magnetization loops $M(H)$ measured perpendicular to the film at 5 K after field-cooling (FC) under 1 T and zero-field-cooling (ZFC) the film. (b) Initial striped domain pattern measured right after ZFC. (c) At H = +650 mT on the ascending branch of the magnetization loop, the white domains have shrunk or disappeared. (d) At H = −400 mT on the descending branch of the magnetization loop after saturating, the pattern is exactly reverse of the initial pattern in (a). Extracted from Kappenberger et al. [29].

### 2.2. Coherent x-ray resonant magnetic scattering

Complementary to MFM, the technique of coherent x-ray magnetic scattering [30, 31] is a powerful tool to study MDM. Under coherent illumination, the material produces a speckled scattering pattern that is a unique fingerprint of the charge and magnetic configuration of the material [32, 33], as illustrated in **Figure 5**. Because x-rays are insensitive to magnetic fields, an in-situ magnetic field of any value and any direction can be applied to the material while collecting the x-ray scattering signal. This allows the study of MDM throughout the entire magnetization process. Also, if mounted on a cryogenic holder, the material can be cooled down under various field cooling conditions, and x-ray scattering can be measured at low temperature. Lastly, the region illuminated by the x-rays is set by the size of the spatial filter used to obtain transverse coherence. The illumination spot size, typically in the order of 20–50 μm, allows covering a relatively large number of magnetic domains, thus providing useful statistical information [34].

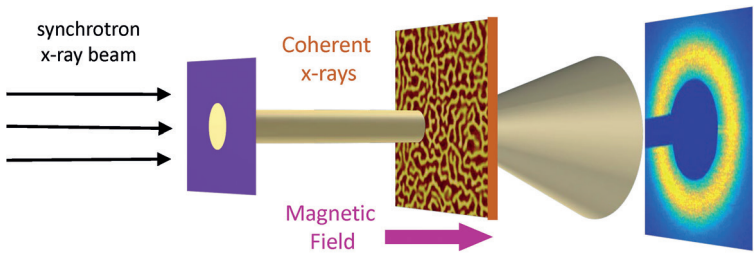

**Figure 5.** Layout for coherent x-ray resonant magnetic scattering (C-XRMS). The synchrotron x-ray beam is spatially filtered by a pinhole to enhance the transvers coherence. The energy of the x-ray is finely tuned to a magnetic resonance, thus providing longitudinal coherence and magneto-optical contrast. Under coherent illumination, the magnetic domains in the film produce a speckle pattern collected on a CCD camera. Extracted from Chesnel et al. [33].

A magneto-optical contrast can be obtained by finely tuning the x-ray energy to specific resonance edges associated to the magnetic elements present in the material. This technique, called x-ray resonant magnetic scattering (XRMS), offers chemical selectivity [35–37]. Typical resonant edges used in XRMS are the $L_2$ and $L_3$ edges which correspond to the excitation of electrons from the (2p) to the (3d) electronic bands in transition metals. For iron (Fe), the $L_{2,3}$ edges are at around 708 and 720 eV, respectively. For cobalt (Co), the $L_{2,3}$ edges are at around 778 and 791 eV, respectively. At these energies (soft x-rays), the x-ray wavelength is around 1–2 nm, thus allowing the probing of magnetic structures in ferromagnetic films, with spatial resolutions down to few nanometers.

By collecting XRMS patterns at various magnetic field values, one can follow the evolution of the magnetic configuration of the material throughout the magnetization process. An example of such experiment is shown in **Figure 6**, where XRMS patterns collected on the descending branch of the magnetization loop are compared to MFM images at nearly same field values. At saturation, no magnetic domain exists so the XRMS pattern has no magnetic signal, only a pure charge background. Shortly after nucleation, sparse domains have nucleated and started to expand, leading to a diffuse disk around the center of the XRMS pattern. As the domain propagation progresses, interlaced domains of opposite directions fill the entire space, leading to a ring-like XRMS pattern. The radius of the ring relates to the magnetic period existing in the material at that stage [38].

Combining coherent x-rays with XRMS, the technique of *coherent*-XRMS (C-XRMS) turns out to be very useful to study MDM [33]. Measured in the scattering space, or the so-called reciprocal space, the CXRMS pattern provides an indirect view of magnetic domains. The inversion of the CXRMS pattern into the real space image is a complex process due to a phase loss in the intensity of the scattering signal [32]. However, if the x-ray beam is made coherent, the x-ray scattering pattern produced by the material will show speckles, as illustrated in **Figure 7**. The specific speckle pattern is a unique fingerprint of the magnetic domain pattern in the real space. Comparing speckle patterns collected at different points throughout magnetization then allows the evaluation of MDM in the material.

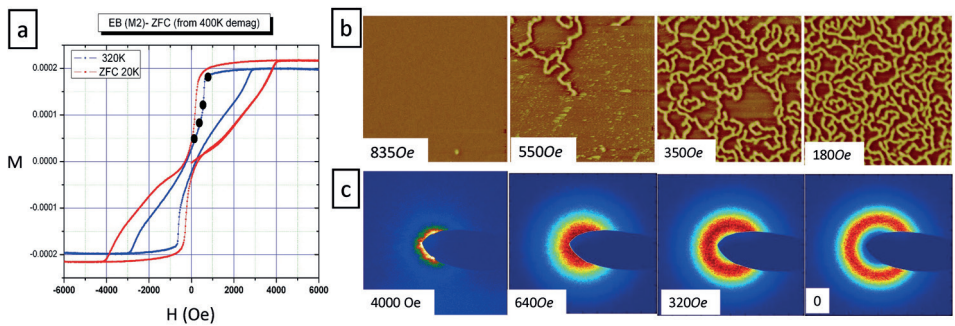

**Figure 6.** Probing magnetic changes in a [Co/Pd]IrMn film via MFM and XRMS. (a) Magnetization loops measured at 320 K and at 20 K after ZFC. (b) Evolution of the magnetic domain pattern at 300 K, viewed via 10 × 10 μm images. (c) Evolution of the XRMS pattern at 20 K in ZFC state at similar points in magnetization. Extracted from Chesnel et al. [38].

**2.3. Speckle cross-correlation metrology**

To evaluate MDM via CXRMS, speckle patterns collected at different field values are compared. This comparison is typically done via cross-correlation [39, 40]. In the cross-correlation process, the intensity in the two images is compared pixel by pixel. In practice, this done by multiplying the intensity of two image, pixel by pixel, while one image is spatially shifted in respect to the other one. Mathematically, this operation may be written as follows:

$$A \times B = \sum_{i,j} A(i,j) B(X+i, Y+j) \qquad (1)$$

where $A$ and $B$ are the two images being correlated, $A(i, j)$ represents the intensity of image $A$ at pixel $(i,j)$ and $B(X+i, Y+j)$ represents the intensity of image $B$ at pixel $(X+i, Y+j)$. Note that a shift of $X$ pixels in the first direction and $Y$ pixels in the second direction is applied between the two images. The sum takes care of covering all the pixels present in the images. The resulting cross-correlation product $A \times B$ is a function of the shift $(X,Y)$. It forms a pattern in the (X,Y) space, which is called correlation pattern. An example of such correlation pattern is shown in **Figure 7e**. To accelerate the cross-correlation operation, which may be computationally costly when comparing thousands of scattering images, the cross-correlation may be done via fast Fourier Transform (FFT) operations. When using FFT, the correlation pattern $A \times B$ is an image of same size than $A$ and $B$. It generally shows a peak around its center, for which X = 0 and Y = 0. The intensity under the peak provides the amount of correlation between the two images. The width of the peak in the (X,Y) space corresponds to the average speckle size, which is generally set by the optics.

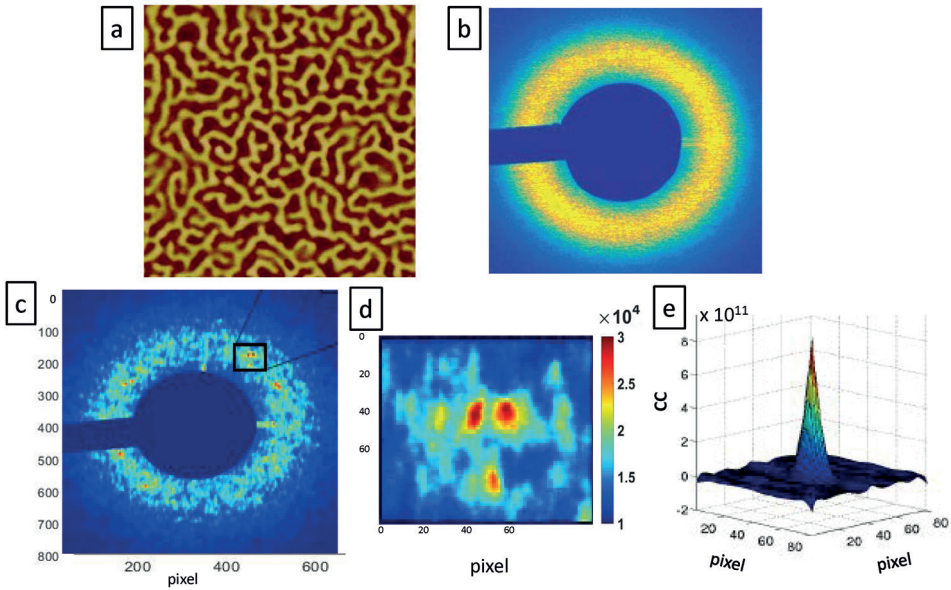

**Figure 7.** Extraction of speckle patterns and cross-correlation process. (a) Magnetic domain pattern in real space, measured via MFM. The image is 10 × 10μm. (b) Associated XRMS pattern as collected on the CCD detector. (c) Speckle pattern, or pure coherent component, extracted from the XRMS pattern (b). (d) Zoomed-in view on the speckle spots. (e) Correlation pattern showing a peak at its center. The area under the peak provides an estimate of the amount of correlation between two speckle patterns. Extracted from Chesnel et al. [33].

To quantify MDM, the intensity in the cross-correlation pattern $A \times B$ is integrated and compared to the intensity of the auto-correlated patterns $A \times A$ and $B \times B$. A normalized correlation coefficient is thus evaluated as follows:

$$\rho = \frac{\sum_{X,Y} A \times B}{\sqrt{\left(\sum_{X,Y} A \times A\right)\left(\sum_{X,Y} B \times B\right)}} \qquad (2)$$

The coefficient $\rho$ is then comprised between 0 and 1. If the two images $A$ and $B$ are completely different, $\rho$ will be close to 0. If the two images $A$ and $B$ are exactly the same, $\rho = 1$ (or 100%). Since each of the correlated speckle patterns is a unique fingerprint of the magnetic domain configuration, the correlation coefficient $\rho$ can be used to quantify MDM. When $\rho$ is low, magnetic domain patterns are very different, and there is no or little MDM. When $\rho$ is high (close to 1), magnetic domain patterns are very similar, and MDM is close to 100%.

### 2.4. Mapping MDM

MDM can be evaluated in a number of different ways. The most straightforward way to evaluate MDM is to compare magnetic speckle patterns measured at the same point $H$ in magnetic field, after a full cycle has been completed, as illustrated in **Figure 8a**. This comparison is called returned point memory (RPM) [39]. An alternative way is to compare magnetic speckle patterns measured at opposite field values. In that case, one point is located at field $H$ on the ascending branch of the magnetization loop whereas the other point is located at opposite field $-H$ is on descending branch. This comparison is called conjugate point memory (CPM). RPM and CPM can then be plotted as function of the magnetic field $H$, where $H$ typically varies from negative saturation to positive saturation, on the ascending branch of the magnetization loop.

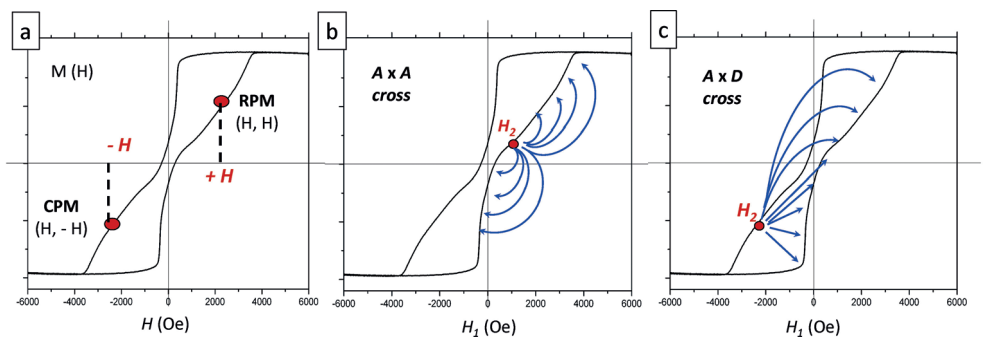

**Figure 8.** Illustration of various cross-correlation approaches. (a) Single point correlation. RPM: cross-correlating pairs of speckle patterns collected at same field (H, H). CPM: cross-correlating pairs of speckle patterns collected at opposite fields (H, −H). (b) Cross-field correlations AxA: correlating any point on the ascending branch with any other point on the ascending branch (c) cross-field correlations AXD: correlating any point on the ascending branch and any other point on the descending branch. In each approach, the two points to be correlated may be separated by several loops. Extracted from Chesnel et al. [38].

A more complete exploration of MDM may be performed by cross-correlating two speckle images $A$ and $B$ collected any field value $H$ throughout the magnetization process [40]. The resulting correlation coefficient $\rho$ may be then mapped in a two-dimensional field space ($H_1$, $H_2$) where $H_1$ is the field value of image $A$ and $H_2$ is the field value of image $B$. The resulting $\rho(H_1, H_2)$ map is called correlation map. The two field values ($H_1$, $H_2$) may be both chosen on ascending branches, either on the same branch or on two different branches separated by one of more cycles. Such correlation map is denoted as A × A. A similar study can be done on the descending branch, leading to a D × D map. If the magnetization loop is symmetrical, the correlations on A × A and D × D are expected to be the same. In another approach, one field value $H_1$ may be chosen on the ascending branch and the other value $H_2$ on the descending branch, leading to an A × D map. Examples of such correlation maps $\rho(H_1, H_2)$ are shown in **Figure 13**.

More subtle measurements of MDM may exploit the spatial information contained in the CXRMS patterns. Instead of computing the cross-correlation on the entire image, the correlation is computed on specific portions of the image. One can for example look at the dependence on the scattering vector $q$, which is the distance from the center of the scattering pattern to a specific point in the image. The correlation is then performed on rings of specific $q$ radii selected from the scattering patterns, as illustrated in **Figure 9**. The resulting correlation coefficient $\rho$ may be plotted as a function of $q$, or mapped in a ($q$, $H$) space [40]. An example of such pattern is shown in **Figure 9d**. Other approaches explore the angular dependency of $\rho$.

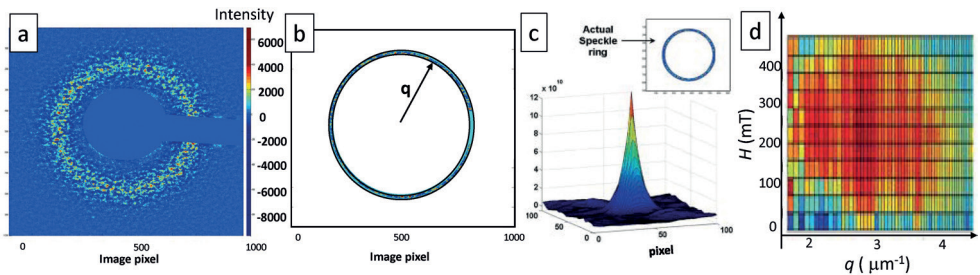

**Figure 9 .** Q-selective correlation process. (a) Initial speckle pattern; (b) ring selection from the speckle pattern; the radius of the ring is Q; (c) correlation pattern resulting from cross-correlating ring selections of two speckle patterns; (d) example of $\rho(Q, H)$ correlation map measure on a ZFC [co/Pd]/IrMn film. Extracted from Chesnel et al. [40].

## 3. Defect-induced MDM

### 3.1. First observations in Co/Pt thin films

The first studies of MDM via x-ray speckle correlation metrology were carried out on [Co/Pt] multilayered thin films by Pierce et al. [39] in 2003. These multilayers, made of 50 repeats of a Co/Pt bilayers, where the Co thickness is 0.4 nm, the Pt thickness is 0.7 nm, exhibit perpendicular anisotropy, leading to the formation of microscopic striped domain patterns.

These magnetic correlation studies demonstrated the occurrence of microscopic magnetic return-point-memory (RPM) when these films exhibited some interfacial roughness. It was found that smooth films with no roughness did not produce any RPM. However, films with large roughness produced significant RPM in the nucleation region of the magnetization process. It was established that the observed microscopic magnetic memory was induced by the presence of defects in the film, playing the role of anchors for magnetic domains to nucleate. This phenomenon is referred to as 'defect-induced' or 'disorder-induced' MDM [41, 42].

In **Figure 10**, magnetic domain configurations are shown for different film roughnesses. When the film is smooth (grown under 3 mT of Ar pressure), a typical interlaced stripe domain pattern occurs. When the film is rough (grown under 12 mT of Ar pressure), the magnetic domain pattern becomes fuzzy and the magnetic periodicity is somewhat lost. The associated XRMS scattering profile shows a well-defined peak for the 3 mT film, but a weaken peak for the 12 mT film.

In these studies, microscopic magnetic correlations were evaluated while cycling the magnetic field throughout minor loops and major loops. In particular RPM and CPM were measured at various field values along the ascending and the descending branches of the major magnetization loop. These measurements were carried out on films with various roughnesses.

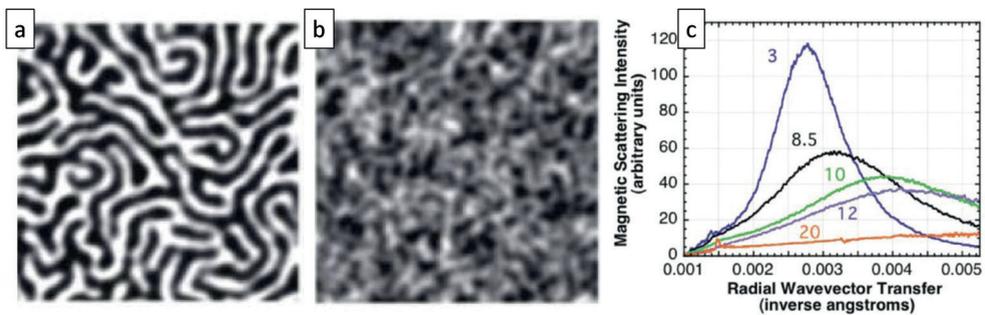

**Figure 10.** Effect of roughness on the magnetic domain configuration in [Co(4 Å)/Pt(7 Å)]$_{50}$ multilayers. (a) and (b) 3 x 3 μm MFM images for films grown under different Ar pressure (a) 3 mT, (b) 12 mT. (c) XRMS profiles for films with different roughnesses (the label indicates the Ar pressure in mT). Extracted from Pierce et al. [41].

### 3.2. RPM and CPM dependence on magnetic field

RPM and CPM were evaluated on the [Co(0.4 nm)/Pt(0.7 nm)]$_{50}$ multilayers at different points throughout the ascending and descending branches of the major loop after cycling the field multiple times. Co/Pt multilayers with significant interfacial roughness exhibited non-zero RPM and CPM, as illustrated in **Figure 11a**. It was found that CPM was systematically lower than RPM. This suggested that disorder has a component that breaks spin reversal symmetry. It was also found that both RPM and CPM exhibited the same trend: their value was larger at nucleation ($H$ = −1 kOe) and decreased monotonically down to near zero when the magnetic field was increased toward saturation ($H$ = +4 kOe). The high correlation value $\rho$ at nucleation suggested that magnetic domains tend to nucleate at specific locations, which are pinned by the defects. The decreasing trend down to zero correlation when field is increasing suggests that once nucleated, the magnetic domains propagate rather randomly, or non-deterministically, throughout the film, leading to decorrelation [43].

### 3.3. Dependence on roughness

As expected, the observed defect-induced MDM depended on the amount of interfacial roughness exhibited by the [Co(0.4 nm)/Pt(0.7 nm)]$_{50}$ multilayers. If the film was relatively smooth, no memory was found. If, on the contrary, the film was rough enough, some RPM and CPM would be detected. The dependence of MDM with interfacial roughness is illustrated in **Figure 11b**, where RPM and CPM are measured at the coercive point. For roughness below 0.5 nm or so, no RPM and no CPM was observed. When the roughness exceeded 0.5 nm, both RPM and CPM quickly increased and started to plateau at around 0.7 nm of roughness, reaching about 40% for CPM and 50% for RPM. This behavior, measured at room temperature was, to some extent, reproduced by nonzero-temperature numerical simulations based on Ising models [41, 42].

The observation of microscopic magnetic memory induced by defects in [Co/Pt] multilayered thin films opened the door to new perspectives and led to further questions, both on the fundamental and applied levels. In particular, the observed CPM and RPM were somewhat limited in magnitude, not exceeding 50–60% in best cases, and also they occurred in a limited region of the magnetization process, namely the nucleation region.

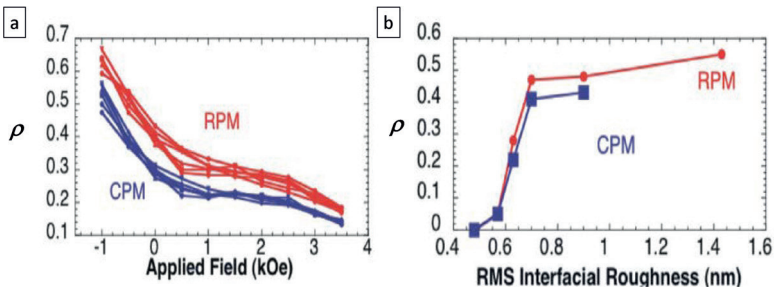

**Figure 11.** RPM and CPM correlations in rough [Co(4 Å)/Pt(7 Å)]$_{50}$ multilayers. (a) RPM and CPM vs. field on the ascending branch of the magnetization loop for a rough film grown under 8.5 mT Ar pressure. (b) RPM and CPM values at the coercive point, plotted as a function of interfacial roughness. Extracted from Pierce et al. [41].

Would there be ways, in some materials, to increase MDM to higher values and to extend it to a wider region of the magnetization loop?

## 4. Exchange-bias induced MDM

In an attempt to increase the amount of MDM initially observed in rough ferromagnetic materials came the idea of incorporating exchange couplings in the film by interlaying the ferromagnetic (F) layer with an antiferromagnetic (AF) layer that would play the role of a magnetic template.

This was successfully achieved by combining a F Co/Pd multilayer with an AF IrMn layer [44]. After cooling the material below its blocking temperature $T_{B'}$ and inducing exchange couplings (EC), high MDM was observed, which extended throughout almost the entire magnetization cycle. The observed MDM reached unprecedented values as high as 100%, even when the film was smooth.

If the material is cooled down under a non-zero field, the EC interactions produce a net exchange bias (EB) in the magnetization loop. However, because the origin of MDM is microscopic, high MDM can be observed even when zero-field cooling the material (in the absence of field), in which case not net bias exists.

### 4.1. First MDM observations in [Co/Pd]IrMn multilayers

The first observations of EC induced MDM were reported by Chesnel et al. [44] in 2008, in [Co/Pd]/IrMn multilayers. It was found that when zero-field-cooled (ZFC), the material exhibited high MDM. The magnetic correlations reached high values throughout a wide range of field values.

The material consisted of an interlay of F [Co(0.4 nm)/Pd (0.7 nm)]$_{12}$ multilayers with AF layers made of IrMn (2.4 nm) alloy, repeated 4 times. This film exhibited PMA, leading to the formation of serpentine magnetic domains that were about 150–200 nm wide, as seen in **Figure 12a**.

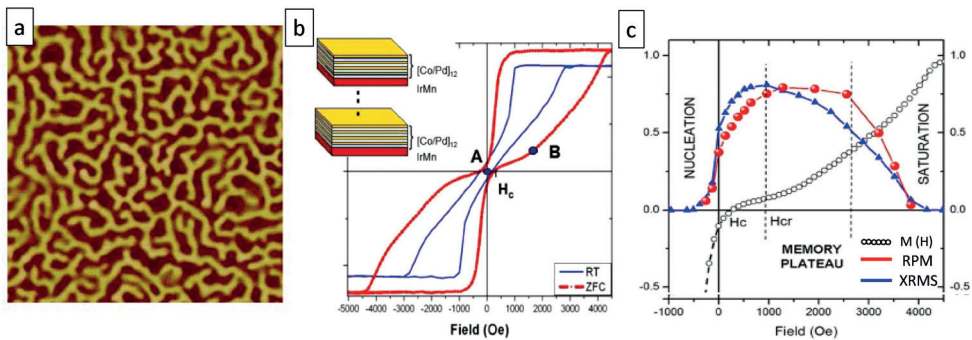

**Figure 12.** MDM measurements in ZFC [Co/Pd]/IrMn multilayers. (a) 10 × 10 μm MFM image of the striped magnetic domains. (b) Magnetization loops $M(H)$ at 300 K and at 20 K in ZFC state. (c) RPM vs. field $H$ plotted against the ascending branch of the magnetization curve. Also plotted is the intensity of the XRMS signal. Extracted from Chesnel et al. [44].

To enhance the exchange couplings between the [Co/Pd] and the IrMn layers, the film was first demagnetized at high temperature and was then cooled down below its blocking temperature, $T_B \sim 300$ K, down to temperatures as low as 20 K. MDM was probed at 20 K using x-ray speckle correlation metrology. CXRMS patterns were collected at finely spaced field values throughout the magnetization loop seen in **Figure 12b** while an in-situ magnetic field was cycled numerous times.

### 4.2. Dependence on magnetic field throughout magnetization loop

The MDM observed in the ZFC [Co/Pd]/IrMn film shows an interesting behavior, which drastically differs from the behavior observed in rough Co/Pt films. **Figure 12c** shows the amount RPM measured in [Co/Pd]/IrMn at 20 K after ZFC and its dependence on magnetic field throughout the ascending branch magnetization loop. Unlike for rough Co/Pt films, RPM is low at nucleation. However, as the magnetic field increases, RPM rapidly increases when the field approaches the remanent coercive point $H_{cr}$ and plateaus at values as high as 80–90%. The plateau extends over a wide field region above the coercive point. RPM eventually decreases down to zero, when the field approaches saturation.

### 4.3. Cross-field MDM maps

In addition to measuring RPM, cross-field magnetic correlations were further carried out throughout the entire magnetization loop. Speckle patterns collected along the ascending branch and along the descending branch of the magnetization loop were cross-correlated. Resulting correlation maps $\rho(H_1, H_2)$ are shown in **Figure 13**, specifically the A × A map (correlations between two ascending branches) and the A × D map (correlations between ascending and descending branches) measured after one field cycle. The RPM and CPM information may be sliced from these map along their diagonal. Both the A × A and A × D correlation maps show high MDM, with $\rho$ reaching as high as 95 ± 5% in the central region where $H_1 \approx H_2 \approx H_{cr}$. The high correlation forms a plateau extending throughout a wide region of field values from near above nucleation to near below saturation. This is not only true along the diagonal, where $H_1 = H_2$, but also off-diagonal where $H_1 \neq H_2$, revealing that the nanoscale magnetic domain patterns formed at these field values are all very similar [38].

### 4.4. Imprinting of a magnetic template via field cooling

The drastic behavioral differences between the MDM observed in ZFC [Co/Pd]/IrMn films and the MDM observed in rough Co/Pt films arise from the different microscopic magnetic interactions. In one case, MDM is induced by defects or disorder. In the other case, MDM is induced by exchange couplings. In this later case, MDM exits even in the absence of defects, that is, in smooth films. In the absence of defects, MDM is low at nucleation, as observed in **Figure 12**, because domains nucleate at random locations in the film. The reason for high MDM to occur at higher field values is the presence of a magnetic template imprinted in the AF layer during the cooling.

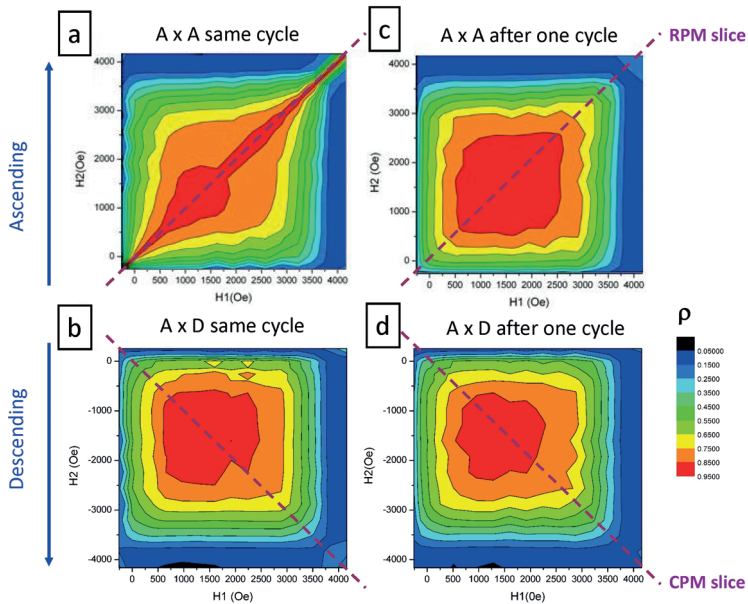

**Figure 13.** Cross-field MDM maps measured on [Co/Pd]/IrMn multilayers at 20 K in ZFC state. (a) A × A map measured on same cycle, (b) A × D map measured on same cycle, (c) A × A map measured after one cycle, (b) A × D map measured after one cycle. Extracted from Chesnel et al. [38].

In the specific material, the magnetic pattern gets imprinted from the F Co/Pd layer into the AF IrMn layer through exchange couplings via interfacial uncompensated spins. In ZFC state, the imprinted pattern is typically formed of interlaced magnetic domains with opposite magnetization, pointing perpendicular to the material, either out-of-plane or into the plane, as illustrated in **Figure 14a, b**. The area covered by one magnetization direction nearly equals the area covered by the other magnetization direction. When, in the ZFC state, the external magnetic field is cycled, the magnetic domains in the F layer successively nucleate, propagate, expand and eventually collapse at saturation. However, due to the frozen underlying magnetic pattern in the AF layer, the domain formation process in the F layer is not random (as it would be for a single smooth F layer). The domain formation process is highly guided by exchange coupling interactions with the underlying frozen template, so that when the remanent coercive point is reached, the domain pattern exactly matches the imprinted one [38, 44].

Guided by the exchange interactions with the magnetic template imprinted in the AF layer, the magnetic domain formation and reversal in the F layer is deterministic. The magnetic domains in the F layer always form in a way to match the underlying template. Consequently, high MDM occurs from about the remanent coercive point $H_{cr}$, where the template is fully matched, all the way up to nearly saturation, as domains evolve in a way to conserve the template topology. This happens both on the ascending and the descending branches of the magnetization loop, as illustrated in **Figure 14c**. Also the correlations between the ascending and descending branches is high. This is possible in the ZFC state because the imprinted pattern is made of long interlaced stripes with equal coverage of up and down domains.

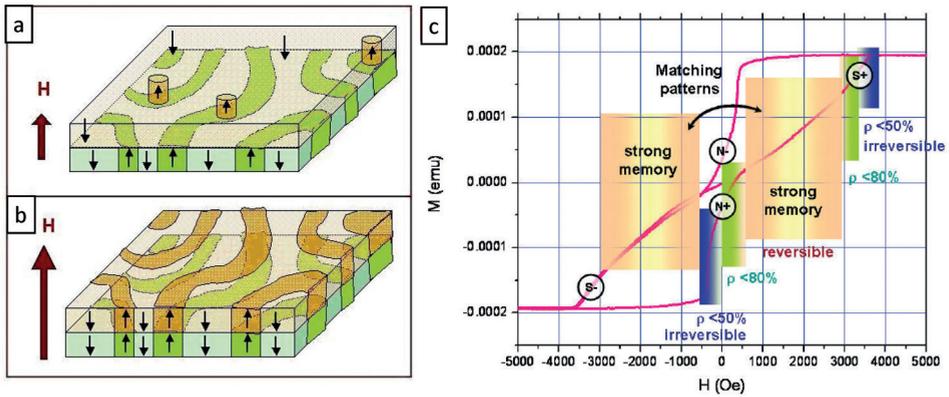

**Figure 14.** Illustration of the origin of MDM in exchange biased [Co/Pd]/IrMn multilayers. (a) and (b) Schematics of the domain pattern forming in the F layer (top) and AF layer (bottom) in a ZFC state where a striped domain pattern is imprinted in the AF layer (a) at nucleation, sparse bubble nucleate randomly in the F layer, (b) at the coercive point, the F domain pattern matches the AF imprinted pattern. (c) Schematics showing the occurrence of MDM throughout the magnetization cycle. MDM is the strongest in the central region of the magnetization loop from the remanent coercive point all the way up to near saturation, both on the ascending and descending branches of the magnetization loop. Extracted from Refs. [38, 44].

## 5. Persistence of MDM through field cycling

When MDM is observed, a question that arises is if the memory persists through multiple field cycles. This question was investigated both in disorder-induced MDM and in exchange-bias induced MDM. In both cases, the magnetic correlations persisted through field cycling.

**5.1. Field cycling dependence measurements**

In the ergodic assumption, the measured correlation, which is an average over a statistical ensemble of microscopic magnetic domains, is expected to be the same than the time averaged magnetic correlation one would measure at any given location of the material. Because the area of the material probed by the x-rays typically includes thousands of magnetic domains and the system is at equilibrium, the ergodic assumption applies. It is therefore expected that the magnetic correlation measured between two magnetic states separated by a given number of cycles should not change in time.

For statistical purposes, correlation coefficients measured at the same number of separating loops but at different times have been averaged. For example, if speckle patterns were collected throughout five subsequent field cycles, the correlation coefficients between cycles

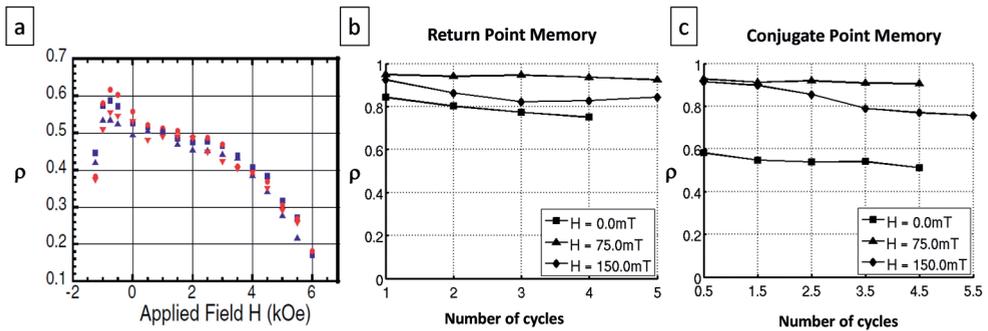

**Figure 15.** Persistence of MDM through field cycling. (a) RPM measured in Co/Pt multilayers after 1 cycle (red) and 11 cycles (blue). (b) and (c) MDM measured in exchange-biased [Co/Pd]/IrMn films (b) RPM vs. field cycle (c) CPM vs. field cycle. Extracted from Refs. [39, 45].

1 & 2, cycles 2 & 3, cycles 3 & 4 and cycles 4 & 5 were all averaged, producing an average correlation coefficient corresponding to one separating cycle. Similar averages were applied for two separating cycles (averaging cycles 1 & 3, 2 & 4 and 3 & 5), three separating cycles, etc. Ultimately, the average correlation coefficient $\rho$ was studied as a function of the number of separating cycles [45].

**5.2. Dependence of MDM on number of field cycles**

RPM and CPM correlations in [Co/Pd] multilayers and in [Co/Pd]/IrMn films were measured throughout many field cycles. In **Figure 15a**, the average RPM measured in Co/Pt films after 1 and after 11 cycles is plotted against the field $H$ on the ascending branch of the magnetization loop. Even after 11 loops, RPM is as high as after one cycle [39]. In **Figure 15b, c**, the average RPM and CPM values measured in ZFC [Co/Pd]/IrMn films at specific points in field are plotted as a function of number of separating cycles. Both RPM and CPM appeared nearly constant as the number of separating cycles was increased. In particular, the optimal RPM value, which occurs near the coercive region, remains within 95–100%, and the optimal CPM in that region remains within 90–95%. The observed persistence of MDM with field cycling is consistent with predictions for exchange-bias induced magnetic memory. Indeed, in these systems, the magnetic domain template imprinted in the AF layer is frozen, meaning it does not change while the magnetic field is cycled, as long as the temperature is kept constant below the blocking point. The domain pattern in the F layer will therefore tend to always retrieve that same unchanged magnetic template, independently from the number of field cycles [45].

## 6. Spatial dependence of EC-induced MDM

The MDM results previously discussed in this chapter were obtained by cross-correlating entire speckle patterns altogether. The associated correlation numbers provided an ensemble average microscopic information, but no spatial dependency was probed. Spatial information is however included in the 2D speckle patterns which are being correlated for the estimation of MDM. This spatial information can be exploited to extract information about possible spatial dependency in MDM. Interesting oscillatory spatial dependence was found in the MDM exhibited by [Co/Pd]IrMn multilayers.

### 6.1. Exploring spatial dependence

Each 2D speckle pattern, such as the one in **Figure 7**, represents the intensity of the x-rays coherently scattered by the material. Because the scattering process involves an inversion from the real space to the scattering space, the spatial scale on the speckle images is an inverse of a distance. A common quantity to locate positions in the speckle patterns is the scattering vector $q$. The origin of the vector $q$ is the center of the scattering pattern, as illustrated in **Figure 9**. The magnitude of $q$ indicates spatial scales $d$ being probed in the real space. $q$ and $d$ are linked by an inverse relationship: $q = \frac{2\pi}{d}$. The larger $q$ is, the smaller the features in the real space. The spatial scales probed when collected scattering patterns such as the one in **Figure 7** typically range from about 50 nm up to a few microns.

Most scattering patterns observed when probing magnetic domain patterns in PMA films show a ring. The presence of the ring reveals a magnetic periodicity in the domain pattern with an isotropic arrangement (no preferred direction). The radius $q^*$ of the ring represents the magnetic period $d^*$. This magnetic period is basically twice the domain width, as one period include a pair of up and down domains. In the [CoPd]/IrMn films, the observed magnetic period was typically $d^* \sim 300$–350 nm.

By cross-correlating selected regions of the scattering speckle pattern rather than the entire pattern, one obtains spatially dependent correlation numbers. This is useful to detect any spatial correlation features occurring at the nanoscale. For instance, magnetic domains patterns may correlate well at the short scale (100–500 nm), but not correlate so well in the long range (1 μm and above) or vice versa.

### 6.2. Space-field maps of MDM

To study the spatial behavior of MDM, in particular its dependence on $q$, cross-correlation is performed on selected rings of the speckle patterns. The rings are concentric, centered about the origin of the scattering pattern. The radius $q$ of the ring is increased from zero to the largest size accessible in the image, thus providing a spatially dependent correlation coefficient $\rho(q)$ [40].

Because speckle patterns are collected at specific points in field $H$ along the magnetization loop, the exploration of spatial dependence in MDM may be done at each field value $H$. Ultimately, the correlation $\rho$ is mapped in a two-dimensional ($q$, $H$) space, thus probing the evolution of the spatial dependence in MDM with field. An example of such $\rho(q, H)$ is shown in **Figure 16a**.

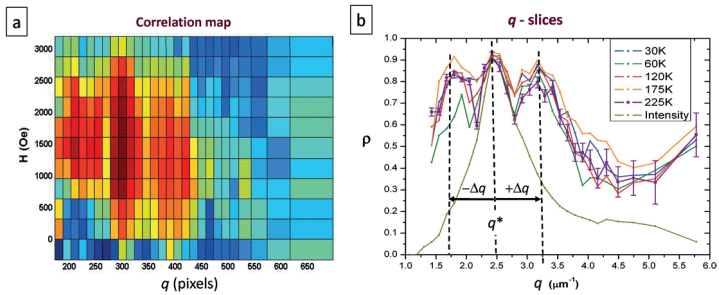

**Figure 16.** Q-dependence of MDM in ZFC exchange-biased [Co/Pd]/IrMn films. (a) $\rho(Q, H)$ correlation map measured at 30 K. (b) $\rho(q, H_c)$ slices through the correlation maps at $H = H_c$ measured at different temperatures from 30 K up to 225 K, showing an oscillation at all temperatures. Extracted from Chesnel et al. [46].

## 6.3. Oscillatory behavior of MDM in [Co/Pd] IrMn

The spatial and field dependence of MDM was explored in [Co/Pd] IrMn films [46]. The films were ZFC below the blocking temperature. Speckle patterns were collected at low temperature, throughout the magnetization cycle and several subsequent loops. Ring selective cross-correlations were carried out. The resulting $\rho(q, H)$ maps, averaged over subsequent cycles, showed interesting features. In particular, slices through the maps at specific $H$ values around the remanent coercive point $H_{cr}$ showed an oscillatory behavior for $\rho(q)$, as seen in **Figure 16b**.

The oscillation observed in the $\rho(q, H_c)$ curve in the ZFC [Co/Pd] IrMn films revealed spatially dependent MDM. The central peak in the $\rho(q, H_c)$ curve seen on **Figure 16b**, occurred at the same location $q^*$ than the ring in the scattering pattern. This suggested that MDM is strongly correlated with the magnetic period in the magnetic domain pattern, here around 400 nm. Two satellite peaks were observed around the central peak, at the same distance $\pm \Delta q$ from $q^*$. The presence of these satellite peaks suggested a spatial superstructure in MDM. The size of the superstructure, set by the value $\Delta q$, was found to be about $D \sim 1.5$ μm. Topographical AFM images of the surface of the films indicated that $D$ nearly matched the average distance between structural defects in the material.

## 6.4. Azimuthal angular dependence

Since the speckle patterns used for the evaluation of MDM are 2D images, one can explore the spatial dependence of MDM in at least two directions. In addition to probing the dependence on the scattering vector $q$, one can investigate the angular dependence, while varying the azimuthal angle $\Delta$ that indicates a particular location on the ring [47], as illustrated in **Figure 17**. In that case, cross-correlations are measured between two points on the ring separated by an angle $\Delta$, leading to a correlation coefficient $\rho(q, \Delta)$. Such study was carried out on ZFC [Co/Pd] IrMn films and showed some periodical variations as a function of the angle $\Delta$ as illustrated in **Figure 17c**. These observations suggested the existence of some hidden rotational symmetries in the formation of the disordered magnetic domain patterns in these films.

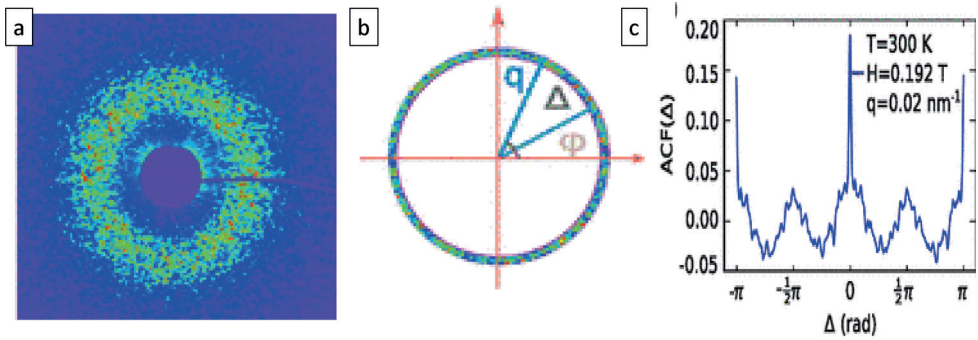

**Figure 17.** Angular dependence of MDM in ZFC exchange-biased [Co/Pd]/IrMn films. (a) Speckle pattern collected near the coercive point. (b) Selected ring from the speckle pattern (a). The angles $\theta$ and $\Delta$ are defined here. (c) Autocorrelation function plotted against $\Delta$. Extracted from Su et al. [47].

## 7. Dependence of EC-induced MDM on temperature

The results discussed in the previous sections show the occurrence of MDM via exchange bias when the film is cooled down to low temperatures, well below the blocking temperature $T_B$. Most of the measurements were performed at around 20 K after ZFC cooling the material. A question that arises next is if the EC-induced MDM depends on temperature and how it behaves at the phase transition near $T_B$. These questions were investigated in exchange biased [Co/Pd]IrMn films [38].

### 7.1. Probing temperature dependence in zero-field-cooled (ZFC) state

To probe the temperature dependence of MDM, the [Co/Pd]/IrMn films were first heated up to around 400 K, well above the blocking temperature $T_B \sim 300$ K and demagnetized at 400 K. The films were then ZFC down to 20 K. Speckle patterns throughout many magnetization cycles were collected in the ZFC state at 20 K, and then at higher temperatures while warming the film all the way back up to above $T_B$. Average A × A and A × D correlation maps were measured at several points in temperature. For each point, the temperature was finely controlled and stabilized via a cryogenic environment. The resulting correlation maps are shown in **Figure 18**.

### 7.2. Persistence of MDM through warming up

The correlation maps in **Figure 18a** show the occurrence of high MDM extending throughout a wide region of the magnetization loop at all temperatures below $T_B$. The high MDM occurred for both A × A and A × D correlation maps between same branches or opposite branches of the magnetization loop, respectively. At all temperatures below $T_B$, MDM reached a high value plateauing over a large region of field values around the coercive point.

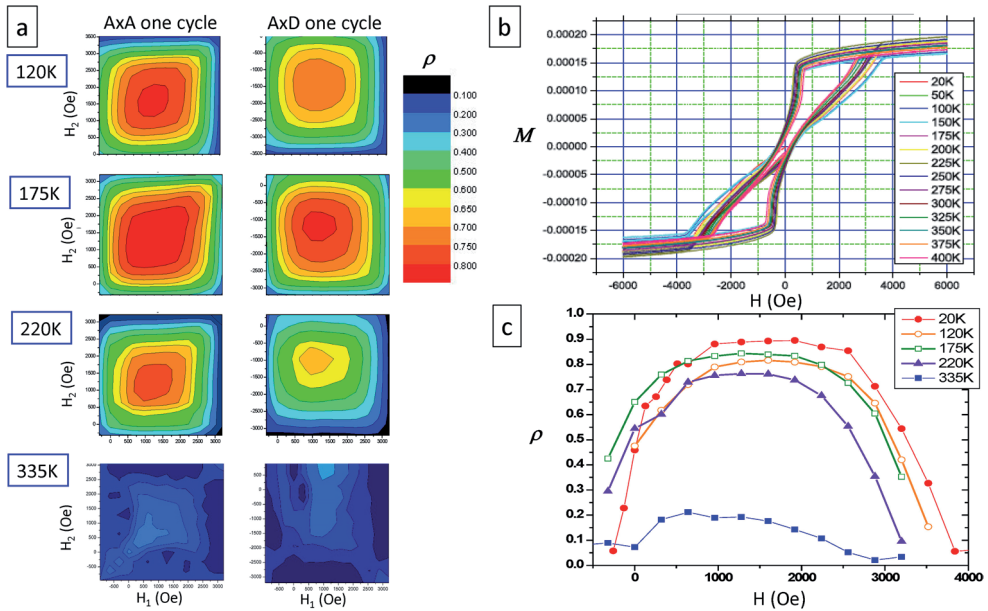

**Figure 18.** Temperature dependence of MDM in ZFC exchange-biased [Co/Pd]/IrMn films. (a) A × A and A × D correlation maps measured after one cycle at different temperatures while warming from 20 K up to 335 K. (b) Magnetization loop at various temperatures. (c) Slices through the A × A maps at $H \approx H_{cr}$. Extracted from Chesnel et al. [38].

The temperature dependence of MDM may be compared to the temperature dependence of the magnetization loop. The shape of the magnetization loop in the ZFC state, as seen in **Figure 18b** is symmetrical, centered about the origin. The loop has an hourglass-like shape; it is narrow at the center and opens up at the extremities, due to the presence of exchange couplings. When increasing the temperature from the ZFC state, the overall shape of the magnetization loop remains the same, but the magnitude of the opening, or hysteresis, progressively decreases. Despite the changes observed in the magnetization loop while warming the film up from the ZFC state, MDM remains strong and extended at all temperatures below $T_B$. Slices through the correlation maps, shown in **Figure 18c**, all show the same trend: low correlation at nucleation, sharply increasing to reach a high correlation plateau in the central region of the magnetization loop, and then sharply decreasing toward saturation.

### 7.3. Disappearance of MDM above the blocking temperature

When the temperature is increased above $T_B$, MDM vanishes, as shown in **Figure 18**. At that stage, the correlation map shows nearly zero correlation, becoming all blue.

The persistence of high MDM throughout warming, at all temperatures below $T_B$, and its vanishing above $T_B$ confirms that MDM is here purely induced by exchange-couplings between the F and AF layers. Above $T_B$, these exchange couplings disappear. Consequently, the magnetic domain template which was imprinted in the AF layer throughout the uncompensated spins at the interface is lost once the magnetic field is cycled again. In the absence of magnetic template, and since the film is smooth (absence of defects), the magnetic domains in the F layer nucleate and propagate randomly.

## 8. Optimizing EC-induced MDM by adjusting field cooling conditions

All the studies of EC-induced MDM discussed in the previous sections were carried out on zero-field-cooled [Co/Pd]/IrMn films. Next question is what happens if the film is now field-cooled (FC), i.e., cooled in the presence of a magnetic film. Will MDM remain as high in the FC state as it is in the ZFC state?

This question was investigated by studying the magnetic correlations after cooling the [Co/Pd]/IrMn films under various cooling field conditions. The resulting correlation maps showed drastic changes as a function of the magnitude of the cooling field [33]. The result of this study is summarized in **Figure 19**. Magnetic correlations were measured in [Co/Pd]/IrMn films after field cooling the material under magnetic field of various magnitudes $H_{FC}$, increasing from zero (ZFC state) all the way up to high values, near saturation value $H_s$.

### 8.1. Dependence of MDM on cooling field magnitude

When the magnitude $H_{FC}$ of the cooling field is near zero, high MDM, up to 100%, is observed. This high MDM extends from nucleation (lower left corner of the correlation map) all the way up to saturation (upper right corner of the correlation map).

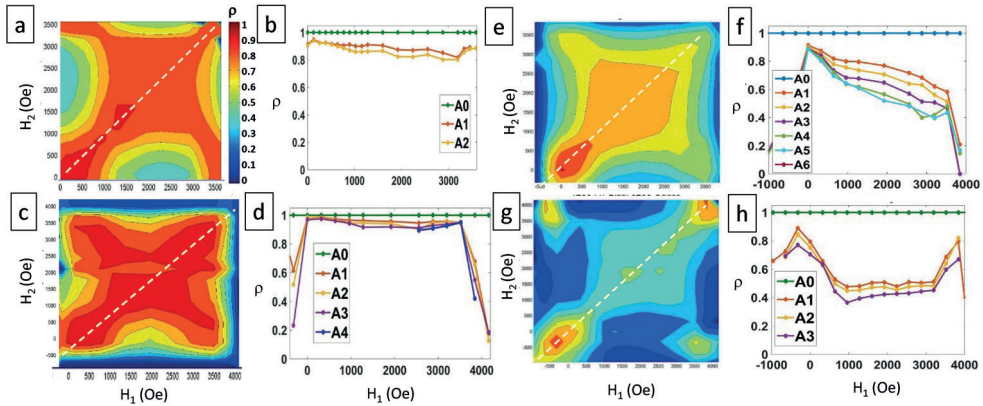

**Figure 19.** Cooling condition dependence of MDM in [Co/Pd]/IrMn films. (a) A × A correlation map measured at 20 K after ZFC; (b) slices through map (a); (c) A × A map measured at 20 K after cooling under $H_{FC}$ = 2240 Oe; (d) slices through map (c); (e) A × A map measured at 20 K after cooling under $H_{FC}$ = 2560 Oe; (f) slices through map (e); (g) A × A map measured at 20 K after cooling under $H_{FC}$ = 3200 Oe; (h) slices through map (g). Extracted from Chesnel et al. [33].

When $H_{FC}$ is increased up to moderately high values, high MDM is still observed, with a wide correlation plateau extending throughout a large range of field values from nucleation to saturation.

When $H_{FC}$ approaches the saturation value, which for the [Co/Pd]/IrMn film is around $H_s \approx 3200$ Oe, the measured magnetic correlations rapidly decrease. At $H_{FC} \approx 2600$ Oe, the correlation drops from about 90% at nucleation down to about 50% in the central region of the magnetization loop.

When $H_{FC} \approx H_s \approx 3200$ Oe, MDM has almost vanished, except at the nucleation and saturation extremities of the magnetization loop.

**8.2. Shaping MDM by adjusting cooling field conditions**

In the ZFC state, the magnetic domain template imprinted in the AF layer is formed of long interlaced stripes with about the same amount of domains of opposite magnetizations (about a 50% up: 50% down split). The imprinted domain pattern constitutes a template for the domain reversal in the F layer. Because of exchange couplings occurring between the Co spins in the F layer and the uncompensated interfacial Mn spins in the AF layer, the domains form in a way to match the underlying template. The matching occurs rapidly when approaching the coercive point (where the up-down magnetization split is exactly 50–50%). The matching persists at higher field values, all the way up to near saturation. This is possible because, even though the magnetic domains of opposite magnetization expand and shrink, the specific topology of the domain pattern still matches the underlying template, as illustrated in **Figure 20**.

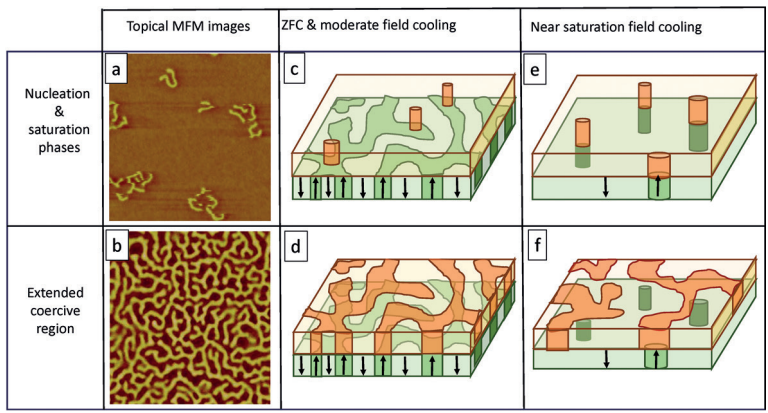

**Figure 20.** Illustration of the magnetic domain memory effect in the exchange-biased [Co/Pd]/IrMn film under various field cooling conditions. The imprinted pattern in the AF IrMn layer is shown in green color, the domain pattern in the F layer is shown in orange color. Arrows indicate magnetization direction. (a, b) 5 × 5 μm MFM images showing actual magnetic domains in the F Co/Pd layer at room temperature: (a) near nucleation and (b) near the coercive point. (c, d) Sketches of the magnetic domain configuration for ZFC and moderate field-cooled states: (c) near nucleation and (d) near the coercive point. (e, f) Sketches of the magnetic domain configuration in near-saturating field-cooled states: (e) near nucleation and (f) near the coercive point. Extracted from Chesnel et al. [33].

In the FC states, the magnetic domain template imprinted in the film has an unbalance of up and down domains, and the net magnetization is non-zero. However, if the magnitude $H_{FC}$ of the cooling field remains in a certain range above $H_{cr}$, the imprinted pattern, still formed of interlaced up and down domains, provides a template that entirely drives the reversal of the magnetic domains in the F layer. The topology of the imprinted pattern in the FC state resembles that of the imprinted pattern in the ZFC state. High MDM is therefore maintained throughout the entire magnetization process.

If the magnitude $H_{FC}$ of the cooling field approaches saturation value $H_s$, the imprinted magnetic domain pattern does not include long interlaced magnetic domains of opposite magnetization anymore but a few bubble domains sparsely scattered throughout the film. This imprinted template is not able to drive the magnetic domain formation throughout the entire reversal but only at the extremities, that are the nucleation and the saturation points, as illustrated in **Figure 20**.

These results demonstrate the possibility to induce and control nanoscale MDM in exchange biased films by adjusting the field cooling conditions. High, up to 100% MDM, extending throughout the entirety of the magnetization loop can be achieved by cooled under no field or relatively low field values. However, MDM can be almost eliminated, by cooling the material under high magnetic field, approaching saturation and higher.

## 9. Conclusion

Magnetic domain memory (MDM) is an unusual property exhibited by certain ferromagnetic films, where the microscopic magnetic domains tend to reproduce the same topological pattern after it has been erased by an external magnetic field. In most ferromagnetic materials, MDM does not occur. When an external magnetic field is applied and cycled, microscopic magnetic domains form and propagate throughout the material in non-deterministic ways. However, it has been found that some ferromagnetic thin films with perpendicular magnetic anisotropy (PMA) do show significant MDM under certain structural and magnetic conditions. One structural condition is the presence of defects. When rough enough, thin Co/Pt multilayers with PMA, exhibits partial MDM occurring in the nucleation phase of the magnetization process. This disorder-induced MDM is caused by the presence of microscopic structural defects, playing the role of pinning sites for the domain nucleation. Another way to induce MDM and to maximize it, even in smooth films, is incorporating magnetic exchange couplings (EC) between a PMA ferromagnetic (F) film and an underlying antiferromagnetic (AF) film. This has been achieved by combining F [Co/Pd] multilayers with AF IrMn layers. After demagnetizing the material and zero-field-cooling it below its blocking temperature, high MDM up to 100% was observed throughout almost the entirety of the magnetization process. This high MDM is induced by EC interactions between the Co spins in the F layer and the interfacial uncompensated spin in the IrMn layer. When the material is cooled down below its blocking temperature, a specific magnetic domain pattern gets imprinted into the AF layer. When the field is cycled at low temperature, the frozen imprinted AF pattern then plays the role of a template for the domain formation in the F layer. The resulting high MDM extends throughout a wide range of field values, from the coercive point to nearly saturation. This EC-induced MDM persists through field cycling and through warming the material all the way up the blocking temperature, above which MDM vanishes. Additionally, it was found that the amount of EC-induced MDM can be varied by adjusting the magnitude of the field applied during the cooling. If the material is cooled under no field or moderate field, MDM reaches high values up to 100% throughout most of the magnetization process. If, however, the material is cooled under a nearly saturating field, MDM vanishes, except at the nucleation and saturation extremities of the magnetization cycle.

These observations of MDM in certain PMA ferromagnetic films opens the door to more investigations. A particular question is if there are other ways to induce and control MDM in a material. It has been recently found that EC-induced MDM may be affected by light, such as x-rays. If the material is illuminated by too intense x-rays, the film may lose its EC properties and MDM may vanish. This finding, which resonates with the emergent all-optical magnetic switching phenomena observations [48–50], suggest that EC-induced MDM could be controlled by light illumination. Ultimately, the ability to induce and control MDM in PMA ferromagnetic film, either by structural disorder, or by exchange couplings and light illumination, may offer a tremendous potential for improving technological applications in the field of magnetic recording and spintronics.

## Author details

Karine Chesnel

Address all correspondence to: kchesnel@byu.edu

Department of Physics & Astronomy, Brigham Young University (BYU), Provo, UT, USA

64  Magnetism and Magnetic Materials[38] Chesnel K, Wilcken B, Rytting M, Kevan SD, Fullerton EE. Field mapping and temperature dependence of magnetic domain memory induced by exchange couplings. New Journal of Physics. 2013;**15**:023016

[39] Pierce MS, Moore RG, Sorensen LB, Kevan SD, Hellwig O, Fullerton EE, Kortright JB. Quasistatic x-ray speckle metrology of microscopic magnetic return-point memory. Physical Review Letters. 2003;**90**:175502

[40] Chesnel K, Nelson J, Wilcken B, Kevan SD. Mapping spatial and field dependence of magnetic domain memory by soft x-ray speckle metrology. Journal of Synchrotron Radiation. 2012;**19**:293

[41] Pierce MS, Buechler CR, Sorensen LB, Turner JJ, Kevan SD, Jagla EA, Deutsch JM, Mai T, Narayan O, Davies JE, liu K, Dunn JH, Chesnel KM, Kortright JB, Hellwig O, Fullerton EE. Disorder-induced microscopic magnetic memory. Physical Review Letters. 2005;**94**:017202

[42] Pierce MS, Buechler CR, Sorensen LB, Kevan SD, Jagla EA, Deutsch JM, Mai T, Narayan O, Davies JE, Liu K, Zimanyi GT, Katzgraber HG, Hellwig O, Fullerton EE, Fischer P, Kortright JB. Disorder-induced magnetic memory: experiments and theories. Physical Review B. 2007;**75**:144406

[43] Seu KA, Su R, Roy S, Parks D, Shipton E, Fullerton EE, Kevan SD. Microscopic return point memory in Co/Pd multilayer films. New Journal of Physics. 2010;**12**:035009

[44] Chesnel K, Fullerton EE, Carey MJ, Kortright JB, Kevan SD. Magnetic memory in ferromagnetic thin films via exchange coupling. Physical Review B. 2008;**78**:132409

[45] Nelson J, Wilcken B, Chesnel K. Persistence of magnetic domain memory through field cycling in exchange bias thin films. Journal of the Utah Academy of Sciences, Arts & Letters. 2010;**87**:267

[46] Chesnel K, Nelson JA, Kevan SD, Carey MJ, Fullerton EE. Oscillating spatial dependence of domain memory in ferromagnetic films mapped via x-ray speckle correlation. Physical Review B. 2011;**83**:054436

[47] Su R, Seu KA, Parks D, Kan JJ, Fullerton EE, Roy S, Kevan SD. Emergent rotational symmetries in disordered magnetic domain patterns. Physical Review Letters. 2011;**107**:257204

[48] Stanciu CD, Hansteen F, Kimel AV, Kirilyuk A, Tsukamoto A, Itoh A, Rasing T. All-optical magnetic recording with circularly polarized light. Physical Review Letters. 2007;**99**:047601

[49] Mangin S, Gottwald M, C-H Lambert, Steil D, Uhlir V, Pang L, Hehn M, Alebrand S, Cinchetti M, Malinowski G, Fainman Y, Aeschlimann M, Fullerton EE. Engineered materials for all-optical helicity-dependent magnetic switching. Nature Materials. 2014;**13**:286

[50] Shcherbakov M, Vabishchevich PP, Shorokhov AS, Chong KE, Choi D-Y, Staude I, Miroshnichenko AE, Neshev DN, Fedyanin AA, Kivshar YS. Ultrafast all-optical switching with magnetic resonances in nonlinear dielectric nanostructures. Nano Letters. 2015;**15**:6985